%% file: paper.tex
\documentstyle[]{mn}
\input{mnfonts.tex}

\newcommand{\Boomerang}
	{BOOMERanG}

\newcommand{\grb}
	{GRB}

\newcommand{\dm}
	{dark matter}

\newcommand{\mond}
	{MOND}
\newcommand{\Mond}
	{MOND}

\newcommand{\cmb}
        {CMB}

\newcommand{\SDSS}
        {SDSS}

\newcommand{\samethanks}
	{{\Huge $^\star$}}

\newcommand{\elementof}
	{\in}
\newcommand{\twoint}
        {\int \!\!\! \int}
\newcommand{\threeint}
        {\int \!\!\! \int \!\!\! \int}

\newcommand{\vect}[1]
        {\mbox{\boldmath ${#1}$}}

\newcommand{\etal}
	{et al.}
\newcommand{\eg}
	{e.g.}
\newcommand{\cf}
	{cf.}
\newcommand{\ie}
	{i.e.}
\newcommand{\eq}[1]
	{equation~(\ref{equation:#1})}
\newcommand{\eqs}[1]
	{equations~(\ref{equation:#1})}
\newcommand{\Eq}[1]
        {Equation~(\ref{equation:#1})}
\newcommand{\Eqs}[1]
        {Equations~(\ref{equation:#1})}
\newcommand{\sect}[1]
	{Section~\ref{section:#1}}

\newcommand{\sects}[1]
        {Sections~\ref{section:#1}}

\newcommand{\fig}[1]
	{Fig.~\ref{figure:#1}}

\newlength{\singlefigureheight}
\newlength{\doublefigureheight}
\newlength{\triplefigureheight}
\newlength{\squarefigureheight}
\newcommand{\AaA}
        {A\&A}

\newcommand{\AJ}
        {AJ}
\newcommand{\ApJ}
        {ApJ}

\newcommand{\ARAA}
        {ARA\&A}

\newcommand{\MNRAS}
        {MNRAS}

\newcommand{\Nature}
        {Nature}

\newcommand{\omo}
        {\Omega_{\rmn m_0}}

\newcommand{\olo}
        {\Omega_{\Lambda_0}}

\newcommand{\pointmass}
        {point-mass}
\newcommand{\Pointmass}
        {Point-mass}

\newcommand{\mtl}
        {mass-to-light}

\newcommand{\los}
        {line-of-sight}

\newcommand{\loss}
        {lines-of-sight}

\newcommand{\FL}
        {FL}

\newcommand{\gr}
	{GR}
\newcommand{\Gr}
	{GR}

\begin{document}

\title[Lensing in modified dynamics]
{Gravitational lensing in modified Newtonian dynamics}

\author[D.\ J.\ Mortlock and E.\ L.\ Turner]
       {
        Daniel J.\ Mortlock,$^{1,2}$\thanks{
		E-mail: mortlock@ast.cam.ac.uk (DJM);
		elt@astro.princeton. edu (ELT)}
	and Edwin L.\ Turner$^3$\samethanks\ \\
        $^1$Astrophysics Group, Cavendish Laboratory, Madingley Road,
        Cambridge CB3 0HE, U.K. \\
	$^2$Institute of Astronomy, Madingley Road, Cambridge
	CB3 0HA, U.K. \\
	$^3$Princeton University Observatory, Peyton Hall,
	Princeton, NJ 08544, U.S.A. \\
       }

\date{
Accepted. 
Received; in original form 2001 January 22}

\pagerange{\pageref{firstpage}--\pageref{lastpage}}
\pubyear{2001}

\label{firstpage}

\maketitle

\begin{abstract}
Modified Newtonian dynamics (\mond) is an alternative theory of gravity 
that aims to explain large-scale dynamics without recourse
to any form of dark matter. 
However the theory is incomplete, lacking a relativistic counterpart,
and so makes no definite predictions about gravitational lensing. 
The most obvious form that \mond ian lensing might take is that
photons experience twice the deflection of massive particles 
moving at the speed of light, as in general relativity (\gr).
In such a theory there is no general thin-lens approximation
(although one can be made for spherically-symmetric deflectors),
but the three-dimensional acceleration of photons is in the same 
direction as the relativistic acceleration would be.
In regimes where the deflector can reasonably be approximated 
as a single \pointmass\ (specifically low-optical depth 
microlensing and weak galaxy-galaxy lensing), 
this naive formulation is consistent
with observations.
Forthcoming galaxy-galaxy lensing data
and the possibility of cosmological microlensing 
have the potential to distinguish unambiguously 
between \gr\ and \mond.
Some tests can also be performed with extended deflectors, 
for example by using surface brightness measurements of
lens galaxies to model quasar lenses, although the breakdown of the 
thin-lens approximation allows an extra degree of freedom. 
Nonetheless, it seems unlikely that simple ellipsoidal galaxies
can explain both constraints. 
Further, the low-density universe implied by \mond\ must be
completely dominated by the cosmological constant (to fit microwave
background observations), and such models are at odds with the 
low frequency of quasar lenses.
These conflicts might be resolved by a fully consistent relativistic
extension to \mond; the alternative is that \mond\ 
is not an accurate description of the universe.
\end{abstract}


\section{Introduction}
\label{section:intro}

Modified Newtonian dynamics (\mond; Milgrom 1983a) is an
alternative theory of gravity in which either inertia
or the effective gravitational force is changed in the limit of low
accelerations (as opposed to large distances).
Taken in its purest form, a \mond ian universe would 
contain no dark matter, and be dominated by the low
density of baryons generated in primordial nucleosynthesis
(Sanders 1998; McGaugh 1999).
\Mond\ makes definite and generally successful
predictions about the dynamics and properties 
of galaxies (Milgrom 1983b; Mateo 1998; McGaugh \& de Blok 1998;
Sanders 2000), 
groups (Milgrom 1998), clusters (Milgrom 1983c; Sanders 1994)
and large-scale structure (Milgrom 1997).
However \mond\ is not a complete physical theory -- in
particular it has no relativistic counterpart
(Milgrom 1983a; Bekenstein \& Milgrom 1984; Sanders 1997).
Thus there are no concrete cosmological predictions, 
and it is not certain how photons couple to gravitational fields.

Despite the lack of any relativistic theory underpinning \mond, 
it is possible to constrain its properties
by invoking the appropriate Newtonian limits 
and using basic symmetries.
Felten (1984) and Sanders (1998) argued that the universe as a whole is in 
the Newtonian regime and so obeys the standard Friedmann 
equations,\footnote{Thus the cosmology can be specified by the 
usual parameters: the normalised present day
matter density,
$\omo$, the similarly normalised cosmological constant, $\olo$,
and the Hubble constant, $H_0 = 100 h$ km s$^{-1}$ Mpc$^{-1}$.}
and similarly that the early universe is governed 
by general relativity (\gr). 
Taking $H_0 = 70$ km s$^{-1}$ Mpc$^{-1}$,
and assuming a totally baryonic universe,
the nucleosynthesis constraints of 
Tytler \etal\ (2000) imply that $\omo \simeq 0.01$.
McGaugh (2000) showed that the \cmb\ anisotropies measured 
by the MAXIMA-1 (Hannay \etal\ 2001) and 
%Balloon-borne Observations of the Microwave Background and Geophysics
\Boomerang\ (de Bernardis \etal\ 2000) experiments are consistent 
with such a low-density cosmological model provided that the 
universe is spatially flat, \ie, $\olo = 1 - \omo \simeq 0.99$. 
Despite these successes, there are a number of qualitative 
difficulties with a \mond ian cosmology (\eg\ Scott \etal\ 2001),
although some are merely indicative of the great conceptual differences
between \mond\ and more conventional physics.

There has been less focus on gravitational lensing within a
\mond ian framework, primarily as the deflection of light cannot
be explained without recourse to a relativistic theory.
Beckenstein \& Sanders (1994) and Sanders (1997), in attempting 
to derive a relativistic extension of \mond, included gravitational
lensing in their considerations, 
but a more empirical approach is adopted here.
Mortlock \& Turner (2001) used the galaxy-galaxy signal
measured by Fischer \etal\ (2000) to constrain the deflection
law of galaxies to be $A(R) \propto R^{0.1 \pm 0.1}$
over the range  
10 kpc $\la R \la$ 1 Mpc.
This is consistent with the deflection law obtained
by assuming that, as in \gr, photons experience twice the deflection of
massive particles moving at the speed of light
(Qin, Wu \& Zhu 1995).
Thus the asymptotic lensing effect of a \mond ian \pointmass\
matches that of an isothermal sphere, which in turn is 
consistent with most observations of lensing by galaxies
(\eg\ Brainerd, Blandford \& Smail 1996; Kochanek 1996).

Encouraged by the qualitative agreement between data and observations,
the lensing formalism of Qin \etal\ (1995) is explored in more detail
here.
In \sect{model} a number of general results are derived and 
then applied to simple lens models.
These are then compared to observations of 
galaxy-galaxy lensing (\sect{galgal}),
microlensing (\sect{microlens}),
multiply-imaged quasars (\sect{quasars})
and lensing by clusters (\sect{cluster}).
The conclusions are summarised in \sect{concs}.

\section{\Mond}
\label{section:model}

\Mond\ is most generally a modification of
inertia (Milgrom 1983a), but, for gravitationally-dominated
systems, can be more intuitively treated as an increase in 
the gravitational acceleration felt by test particles.
The Newtonian acceleration due to a (static) mass distribution
$\rho(\vect{r})$ is 
%(Newton 16??)
\begin{equation}
\label{equation:newton}
\vect{a}_{\rmn N} (\vect{r}) 
=
- G \threeint_{R^3} 
\rho(\vect{r}) \frac{\vect{r} - \vect{r}^\prime}
{| \vect{r} - \vect{r}^\prime |^3} \, {\rmn d}^3 r^\prime,
\end{equation}
where 
$G$ is Newton's gravitational constant
and
$\vect{r}$ is pos\-i\-tion\footnote{Three-dimensional 
vectors are denoted by lower case (\eg\ $\vect{r}$); two-dimensional
vectors -- projected onto the sky -- are denoted by upper case
(\eg\ $\vect{R}$).}.
The \mond ian acceleration is related to this by 
(Milgrom 1983a)
\begin{equation}
\label{equation:mond}
\vect{a}(\vect{r}) f_{\rmn M}\left[\frac{a(\vect{r})}{a_0}\right] 
= \vect{a}_{\rmn N}(\vect{r}),
\end{equation}
where $a_0$ is the critical acceleration, above which 
the dynamics is close to Newtonian. The exact form of $f_{\rmn M}(x)$ is 
not known, but $f_{\rmn M}(x) = 1$ if $x \gg 1$ 
%[\ie\ $\vect{a}(\vect{r}) = \vect{a}_{\rmn N}(\vect{r})$] 
and 
$f_{\rmn M}(x) = x$ if $x \ll 1$. 
No major \mond ian results are sensitive to 
the form of this function in the regime $x \simeq 1$, and so 
\begin{equation}
\label{equation:f}
f_{\rmn M}(x) = \left\{
\begin{array}{lll}
x, & {\rmn if} & x < 1, \\
& & \\
1, & {\rmn if} & x \geq 1
\end{array}
\right.
\end{equation}
is used here for mathematical simplicity.
It is convenient to invert \eq{mond} so that the acceleration can be
expressed as a function of known quantities. Defining a new function 
$f_{\rmn M}^\prime(x)$ with the same asymptotic limits as $f_{\rmn M}(x)$, 
\mond\ may also be defined by
\begin{equation}
\label{equation:mymond}
\vect{a}(\vect{r}) = {f_{\rmn M}^\prime}^{-1/2} 
\left[ \frac{a_{\rmn N}(\vect{r})}{a_0}\right]
\vect{a}_{\rmn N}(\vect{r}).
\end{equation}
For the particular choice of $f_{\rmn M}(x)$ given in \eq{f}, 
$f_{\rmn M}^\prime(x) = f_{\rmn M}(x)$.

There are some difficulties in understanding 
\eqs{f} and (\ref{equation:mymond})
for complex systems (\eg\ Milgrom 1983a; Scott \etal\ 2001),
but their qualitative intrepretation for a single particle is
reasonably straightforward.
For accelerations much greater
than $a_0$ Newtonian dynamics remains valid, as $\vect{a}(\vect{r})
= \vect{a}_{\rmn N}(\vect{r})$. 
But if $a_{\rmn N}(r) \ll a_0$ 
%(the `deep' \mond ian limit) 
then
$\vect{a}(\vect{r}) = [a_0 a_{\rmn N}(\vect{r})]^{1/2} \hat{\vect{a}}_{\rmn N}
(\vect{r})$ and a test particle would feel an acceleration
in the expected direction, but of a greater magnitude.
The value of the critical acceleration has been determined from
galaxy rotation curves as $a_0 = 1.2 \pm 0.1 \times 10^{-10}$ m s$^{-2}$ 
(Milgrom 1983b;
Begeman, Broeils \& Sanders 1991; McGaugh \& de Blok 1998),
and, modulo the uncertainty in the form of $f_{\rmn M}(x)$,
\mond\ has no other free parameters.
\Eqs{f} and (\ref{equation:mymond}) are sufficient to calculate the
trajectory of a massive particle
(\sect{massbend}), from which a natural formalism for the 
deflection of photons can be extrapolated (\sects{lightbend} 
and \ref{section:lensmodels}). 

\subsection{Deflection of massive particles}
\label{section:massbend}

Within the framework of \mond ian (or Newtonian) dynamics
it is possible to calculate the trajectory of an arbitrary particle.
If, however, the particle is sufficiently light that it does not disturb the 
external mass distribution, and its speed is such that its 
path is nearly linear, its deflection can be approximated 
by integrating the acceleration along the unperturbed 
trajectory (\ie\ a straight line). 
Without loss of generality, this path can be defined as being
parallel to the $z$-axis of a Euclidean coordinate system,
with $\vect{R} = (x, y)$ the two-dimensional impact parameter
(relative to some reference point)
in the plane of the sky.
Thus the deflection angle, $\vect{A} = (A_x, A_y)$,
of a particle moving with speed $v$ in the $z$-direction 
is given by
\begin{eqnarray}
\label{equation:alpha_mass}
A_i(\vect{R}) & \simeq & \frac{v_i}{v} \nonumber \\
& = & \frac{1}{v} 
\int_{- \infty}^{\infty} a_i[\vect{r}(t)]
\, {\rmn d}t \nonumber \\
& = & \int_{- \infty}^{\infty}
{f_{\rmn M}^\prime}^{-1/2} \left\{\frac{a_{\rmn N}[\vect{r}(t)]}{a_0} \right\}
{a_{\rmn N}}_i[\vect{r}(t)]
\, {\rmn d}t ,
\end{eqnarray}
where $i \elementof \{x, y\}$,
and $\vect{a}(\vect{r})$
is defined in \eqs{newton} and (\ref{equation:mymond}).
For a minimally-deflected particle, 
$\vect{r}(t) = [x, y, v(t - t_0)]$, where $t_0$ is an
arbitrary reference time.

%\begin{figure}
%\special{psfile="path.ps" angle=-90 vscale=60 hscale=60
%	hoffset=-15 voffset=28}
%\vspace{\singlefigureheight}
%\caption{The path of a minimally-deflected particle through 
%a gravitational field. The position of the particle at time 
%$t$ is $\vect{r}(t) = [x, y, v (t - t_0)]$ where $\vect{R} = (x, y)$
%is its impact parameter and $v$ its speed. The deflection angle,
%$\vect{A} = (A_x, A_y)$, is assumed to satisfy
%$A \ll 1$, which is almost always the case in observed instances of 
%gravitational lensing.}
%\label{figure:path}
%\end{figure}

\subsection{Deflection of photons}
\label{section:lightbend}

In the standard theory of gravitational lensing
(\eg\ Sch\-neider, Ehlers \& Falco 1992),
the deflection of photons in weak, static gravitational fields
can be calculated from \gr, and is simply 
twice the deflection experienced by a massive particle moving at 
the speed of light, $c$.
This has been confirmed observationally for light grazing the
Sun (Dyson, Eddington \& Davidson 1920; 
Robertson \& Carter 1984) and this simple relationship must 
also be true of \mond ian lensing in the Newtonian limit.
There is no strong evidence that this relationship holds in the 
deep \mond ian limit, but such a model is consistent with 
galaxy-galaxy lensing (Mortlock \& Turner 2001), and this 
hypothesis is adopted henceforth.
The deflection angle of photons can thus be read from \eq{alpha_mass}
as 
\begin{equation}
\label{equation:alpha_light}
A_i(\vect{R}) = 
\frac{2}{c}
\int_{- \infty}^{\infty}
{f_{\rmn M}^\prime}^{-1/2}\left\{\frac{a_{\rmn N}[\vect{r}(t)]}{a_0} \right\}
{a_{\rmn N}}_i[\vect{r}(t)]
\, {\rmn d}t ,
\end{equation}
where again $i \elementof \{x, y\}$ and, for a photon,
$\vect{r}(t) = [x, y, c(t - t_0)]$.

\subsubsection{The thin-lens approximation}
\label{section:thinlens}

In the limit $a_0 \rightarrow 0$, the standard relativistic 
deflection formula can be reproduced by using \eq{newton} to obtain
\begin{equation}
\label{equation:alpha_newton}
{A_{\rmn N}}_i (\vect{R}) = 
\frac{2}{c}
\int_{- \infty}^{\infty}
{a_{\rmn N}}_i[\vect{r}(t)]
\, {\rmn d}t.
\end{equation}
In the limit of small deflection angles (relative to the \los\ extent of
the lens)
the thin-lens approximation 
(\eg\ Schneider \etal\ 1992) can
be invoked. 
\Eq{alpha_newton} can thus be reduced to
\begin{equation}
\label{equation:thinlens}
\vect{A}_{\rmn N}(\vect{R}) = 
- \twoint_{R^2} \frac{4 G \Sigma (\vect{R}^\prime)}{c^2}
\frac{\vect{R}^\prime - \vect{R}}{|\vect{R}^\prime - \vect{R}|^2}
\, {\rmn d}^3 R^\prime,
\end{equation}
where the surface density of the lens is 
\begin{equation}
\Sigma(\vect{R}) = \int_{- \infty}^\infty
\rho(x, y, z) \, {\rmn d}z.
\end{equation}
The application of the thin-lens approximation allows many 
lensing problems to be greatly simplied, and
\eq{thinlens} is
one of the main results of standard gravitational lensing theory.

As given in \eq{alpha_light} the \mond ian deflection angle 
cannot be simplified in general.
In particular,
there is no useful thin-lens approximation:
\eq{alpha_light} cannot be simplified in the same way that
\eq{alpha_newton} can. 
More qualitative arguments imply that the same is likely to be
true of any \mond ian lensing theory (see \sect{rod}).
% see \sects{rod} and \reference{section:twopoint}.
Further, constant \mtl\ ratio models of 
real gravitational lenses cannot reproduce the observed image
configurations
(\eg\ AbdelSalam, Saha \& Williams 1998; \sect{quasarmodel}); this
would almost certainly invalidate 
\mond\ if the thin-lens approximation were valid.
The ambiguity in the \los\ mass distribution of an observed 
deflector is inconvenient (as is the extra integral required 
in the calculation of deflection angles), but could be 
thought of as analogous to the a priori unconstrained 
\dm\ distributions that dominate conventional gravitational 
lensing studies. 

Despite the lack of a general thin-lens approximation, 
some simplifications of \eq{alpha_light} are possible.
In many cases the deflection angle can be decomposed into 
two \mond ian segments (when the photon is far away from the 
lens) and a single Newtonian segment (the major deflection near
the lens). If the deflector is sufficiently localised, 
the latter regime can be handled in the usual way, using the 
thin-lens formalism, and the \mond ian contribution away from the
lens may be approximated by the simple \pointmass\ formula
(\sect{pointmass}), the deflector being `unresolved' at such
great distances. 

\subsubsection{The lens equation}
\label{section:lensequation}

Gravitational lensing can be thought of as a mapping between the 
source plane (an imaginary surface, perpendicular to the \los,
on which a source is located) and the sky or image plane. 
The lens equation
relates the angular position of a source, $\vect{\beta}$,
to the angular position(s) of its image(s), $\vect{\theta}$,
and can be written as (\cf\ Schneider \etal\ 1992)
\begin{equation}
\label{equation:lensequation}
\vect{\beta} = \vect{\theta} + \vect{\alpha}(\vect{\theta}),
\end{equation}
where 
\begin{equation}
\label{equation:A_alpha}
\vect{\alpha}(\vect{\theta}) = \frac{d_{\rmn ds}}{d_{\rmn os}}
\vect{A}\left(d_{\rmn od} \vect{\theta}\right).
\end{equation}

In these definitions, $d_{\rmn od}$, $d_{\rmn os}$
and $d_{\rmn ds}$ are the angular diameter distances between
observer and deflector, observer and source, and deflector and
source, respectively. It is not entirely clear how they vary
with redshift in a \mond ian cosmology.
Even if
the universe obeys the Friedmann equation, as implied by
the results of Sanders (1998), the separation of nearby photons
can only be determined by a relativistic extension of \mond,
although the standard distance measures should be reproduced 
in the limit of large angular separation (\cf\ Linder 1998).
Where values of $d_{\rmn od}$, $d_{\rmn os}$ and $d_{\rmn ds}$
are required, they are calculated for the appropriate low-density
Friedmann model using standard formul\ae\ 
(\eg\ Carroll, Press \& Turner 1992).
This is probably consistent 
with the lensing theory described above, but is no more
than a reasonable assumption.

The lens equation allows the calculation of image positions,
magnifications and distortions.
The magnification of a single image is (Schneider \etal\ 1992)
\begin{equation}
\label{equation:mu_th}
\mu(\vect{\theta}) = \left|
%\det 
\frac{{\rmn d}^2 \vect{\beta}}
{{\rmn d} \vect{\theta}^2}
\right|^{-1},
\end{equation}
and the total magnification of a source is then
\begin{equation}
\label{equation:mu_be}
\mu_{\rmn tot}(\vect{\beta}) = \sum_{i = 1}^{N_i} \mu \left[\vect{\theta}_i
(\vect{\beta})\right],
\end{equation}
where the sum is over the $N_i$ images formed, and 
$\vect{\theta}_i(\vect{\beta})$
is the position of the $i$th image.

\subsubsection{Spherical symmetry}
\label{section:spherical}

If the deflector is spherically symmetric, then 
$\rho(\vect{r}) = \rho(r)$ and \eq{alpha_light} can be written as
\begin{equation}
\label{equation:alpha_light_sph}
\vect{A}(\vect{R}) = - \frac{2 G}{c^2} \vect{R}
\end{equation}
\[
\times
\int_{- \infty}^\infty 
{f_{\rmn M}^\prime}^{-1/2}
\left[
\frac{G M\left(< \sqrt{R^2 + z^2}\right)}{\left(R^2 + z^2\right) a_0}
\right]
\frac{M\left(< \sqrt{R^2 + z^2}\right)}
{\left(R^2 + z^2\right)^{3/2}} \, {\rmn d} z,
\]
where
\begin{equation}
\label{equation:m_inside}
M(< r) = \int_0^r 4 \pi {r^\prime}^2 \rho(r^\prime) \, {\rmn d}r^\prime.
\end{equation}
It is possible 
to write down an analogue of the usual lens equation by expressing
$M(< r)$ in terms of the projected surface density.
Using \eq{m_inside} and expressing $\rho(r)$ in terms of 
an Abel integral (\eg\ Binney \& Tremaine 1987),
\begin{eqnarray}
\label{equation:m_inside_proj}
M(< r) & = & \int_0^r 4 \pi {r^\prime}^2 
\int_{r^\prime}^\infty - \frac{1}{\pi}
\frac{{\rmn d} \Sigma}{{\rmn d} R}
\frac{1}{\sqrt{R^2 - {r^\prime}^2}} \, {\rmn d} R
\, {\rmn d} r^\prime, \nonumber \\
& = & 2
\!\! \int_r^\infty
\!\!
\left[
\frac{r}{R} \sqrt{1 - \frac{r^2}{R^2}}
-
\arcsin\left(\frac{r}{R}\right)
\right]
\!\!
R^2
\frac{{\rmn d} \Sigma}{{\rmn d} R}
\, {\rmn d} R \nonumber \\
& - & \pi \int_0^r R^2 \frac{{\rmn d} \Sigma}{{\rmn d} R} 
\, {\rmn d} R,
\end{eqnarray}
where the order of integration is reversed, making the inner 
integral trivial
(\cf\ Kovner 1987).
This formula is not particularly useful for unbounded 
mass distributions like the isothermal sphere (\sect{iso}),
as its convergence properties are poor.
However in the case of \mond\ galactic mass distributions 
fall off as fast as their luminosity density -- exponentially,
in general -- and so \eq{m_inside_proj} is well behaved,
as well as relating the observed surface brightness of a deflector
directly
to its lensing properties.

The combination of \eqs{alpha_light_sph} and (\ref{equation:m_inside_proj})
are suggestive of an alternative formalism for spherical lenses. 
\Mond ian lensing can be cast in terms of the conventional 
formalism of Schneider \etal\ (1992) if an `effective' surface 
density is defined. Specifying it by either $\rho_{\rmn N}(r)$ or
$\Sigma_{\rmn N} (R)$ (as it
is only relevant in the symmetric case), it is the density profile 
that would, under the assumption of Newtonian mechanics,
produce both the same rotation curve and the same 
deflection law as the model profile produces in \mond.
Given that a \pointmass\ has an asymptotically constant rotation speed
of $v_{\rmn c} = (G M a_0)^{1/4}$
in \mond\ (Milgrom 1983b), the addition of an isothermal component
is suggested, and defining
\begin{equation}
\rho_{\rmn N}(r) = \rho(r) + \frac{M^{1/2} a_0^{1/2}}{4 \pi G^{1/2} r^2}
\end{equation}
is a reasonable approximation to many \mond ian deflectors of true mass $M$.
The same rotation curves and deflection laws would result from
defining $f_{\rmn M}(x) = 1 + [1 + (1 + 4x)^{1/2}] / (2 x)$.

In the limit of $f_{\rmn M}^\prime(x) \rightarrow 1$ and $a_0 \rightarrow 0$,
\eq{alpha_light_sph} reduces to the conventional result that
(\eg\ Schneider \etal\ 1992)
\begin{equation}
\vect{A}_{\rmn N}(\vect{R}) = 
- \frac{4 G}{c^2} \frac{M(<R)}{R^2} \vect{R},
\end{equation}
where 
\begin{equation}
M(<R) = \int_0^r 2 \pi R^\prime \Sigma(R^\prime) \, {\rmn d} R^\prime.
\end{equation}

\subsection{Lens models}
\label{section:lensmodels}

Having developed a plausible formalism for gravitational lensing
within the framework of \mond, more specific results can be 
derived for some simple mass distributions.

\subsubsection{\Pointmass}
\label{section:pointmass}

\begin{figure}
\includegraphics{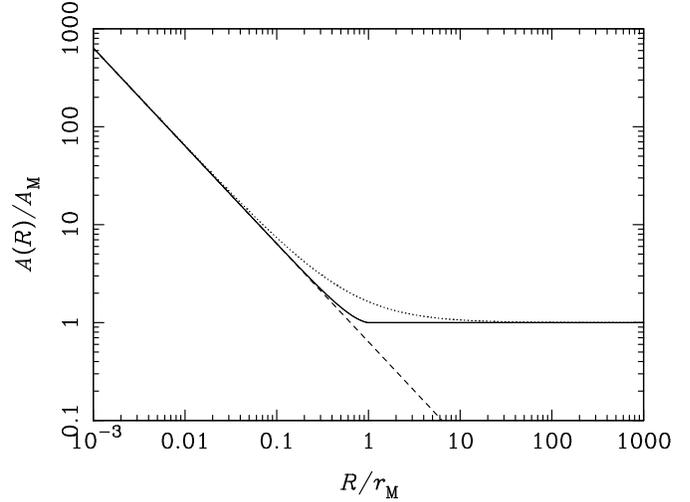}
\vspace{\singlefigureheight}
\caption{The deflection law of a \pointmass\ 
in \mond\ (solid line) and \gr\ (dashed line).
The dotted line shows 
the approximation to the \mond ian result
obtained by calculating the Newtonian deflection
caused by an isothermal sphere superimposed on a \pointmass\
(see \sect{spherical}).}
%The axes are normalised by the radius at which the acceleration
%becomes \mond ian, $r_{\rmn M}$, and the asymptotically constant
%\mond ian deflection angle, $A_{\rmn M}$.}
\label{figure:alpha_sch}
\end{figure}

For a point lens of mass $M$, $\rho(\vect{R}) = M \delta^3(\vect{R})$
and $M(< R) = M$. 
Using the piecewise definition of $f_{\rmn M}^\prime (x)$, 
\eq{alpha_light_sph} splits up into several trivial integrals, 
and the deflection law for a \pointmass\ is given by
\begin{equation}
\label{equation:alpha_sch}
A(R) = 
\end{equation}
\[
\left\{
\begin{array}{lll}
- \frac{4 G M}{c^2 R} \sqrt{1 - \frac{R^2}{r_{\rmn M}^2}}
& & \\
\mbox{}
- A_{\rmn M}
\left[
1 - \frac{2}{\pi} \arctan \left( \sqrt{\frac{r_{\rmn M}^2}{R^2} - 1} \right)
\right],
& {\rmn if} & R \leq r_{\rmn M}, \\
& & \\
- A_{\rmn M} & {\rmn if} &
R > r_{\rmn M},
\end{array}
\right.
\]
where $r_{\rmn M} = (G M / a_0)^{1/2}$ is the distance from a 
\pointmass\ at which the physics changes from the Newtonian to
the \mond ian regime and 
$A_{\rmn M} = 2 \pi (G M a_0)^{1/2} / c^2$
is the asymptotic \mond ian deflection angle. 
This is compared to the standard \pointmass\ deflection angle
(as inferred from the Schwarzschild metric) in \fig{alpha_sch}.
%(which is analogous to Figure 1 of Qin \etal\ 1995)
For large impact parameters ($R \ga r_{\rmn M}$) the photon
experiences an acceleration of less than $a_0$, and so 
the deflection angle is independent of impact parameter, 
in much the same way that the rotation speed would be.
For $R \la r_{\rmn M}$, the light path is in the \mond ian
regime for the most part (the second term in the above formula), 
but the deflection is dominated by 
the Newtonian force close to the lens (the first term, which 
approaches the Schwarzschild bending angle for small $R$). 

The next step in analysing the \pointmass\ lens is to 
convert \eq{alpha_sch} into angular units, in order to write 
down the lens equation. The Schwarschild lens is completely 
characterised by its Einstein angle, 
\begin{equation}
\label{equation:theta_E}
\theta_{\rmn E} = \sqrt{\frac{4 G M}{c^2} 
\frac{d_{\rmn ds}}{d_{\rmn od} d_{\rmn os}}},
\end{equation}
but two other angular scales are useful here:
\begin{equation}
\label{equation:theta_M}
\theta_{\rmn M} = \frac{r_{\rmn M}}{d_{\rmn od}} = 
\sqrt{\frac{G M}{d^2_{\rmn od} a_0}},
\end{equation}
beyond which the deflection angle becomes constant;
and 
\begin{equation}
\alpha_{\rmn M} = \frac{2 \pi \sqrt{G M a_0}}{c^2} 
\frac{d_{\rmn ds}}{d_{\rmn os}} 
= \frac{\pi}{2} \frac{\theta_{\rmn E}^2}{\theta_{\rmn M}}
,
\end{equation}
the value of the deflection angle in this regime.
Combining the above definitions with \eq{A_alpha}, 
\eq{alpha_sch} reduces to
\begin{equation}
\label{equation:alpha_th}
\alpha(\theta) = 
\end{equation}
\[
\left\{
\begin{array}{lll}
- \frac{\theta_{\rmn E}^2}{\theta} \sqrt{1 
- \frac{\theta^2}{\theta_{\rmn M}^2}}
& & \\
\mbox{}
- \alpha_{\rmn M}
\left[
1 - \frac{2}{\pi} \arctan\left( \sqrt{\frac{\theta_{\rmn M}^2}{\theta^2} - 1}
\right)
\right],
& {\rmn if} & \theta \leq \theta_{\rmn M}, \\
& & \\
- \alpha_{\rmn M}, & {\rmn if} & \theta > \theta_{\rmn M},
\end{array}
\right.
\]
with $\alpha(\theta_{\rmn M}) \simeq - \alpha_{\rmn M}$. 
From \eq{mu_th} the magnification of a single image is then
\begin{equation}
\label{equation:mu_point}
\mu(\theta) = 
\end{equation}
\[
\left\{
\!\!\!
\begin{array}{lll}
\multicolumn{3}{l}{
\theta^4 \theta_{\rmn M}^2 /
\left| \left(\theta^2 \theta_{\rmn M} 
+ \theta_{\rmn E}^2 \sqrt{\theta_{\rmn M}^2 - \theta^2} \right) 
\left\{
\theta^2 \theta_{\rmn M} 
- \theta_{\rmn E}^2 \sqrt{\theta_{\rmn M}^2 - \theta^2}
\right.
\right.} \\
\left.
\left.
- \alpha_{\rmn M} \theta \theta_{\rmn M}
\left[ 1 - \frac{2}{\pi} \arctan
\left(\sqrt{\frac{\theta_{\rmn M}^2}{\theta^2} - 1} \right) \right]
\right\} \right|,
& \!\! {\rmn if} \!\! & \theta \leq \theta_{\rmn M}, \\
& & \\
\theta / (\theta - \alpha_{\rmn M}), 
& \!\! {\rmn if} \!\! & \theta > \theta_{\rmn M},
\end{array}
\right.
\]
whereas for a Newtonian \pointmass\ it is simply
$\mu_{\rmn N}(\theta) = |\theta^4 / (\theta^4 - \theta_{\rmn E}^4)|$.

\begin{figure}
\includegraphics{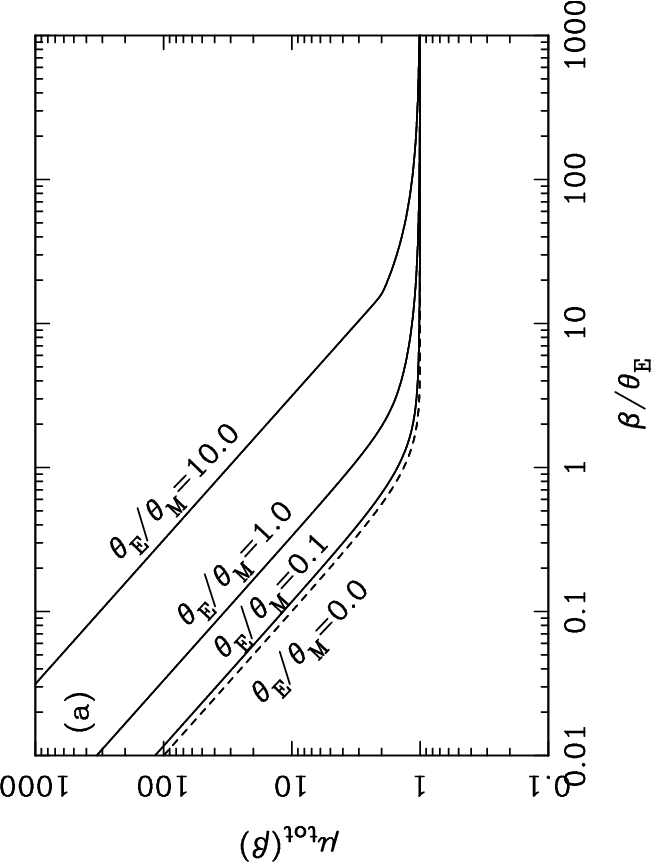}
\includegraphics{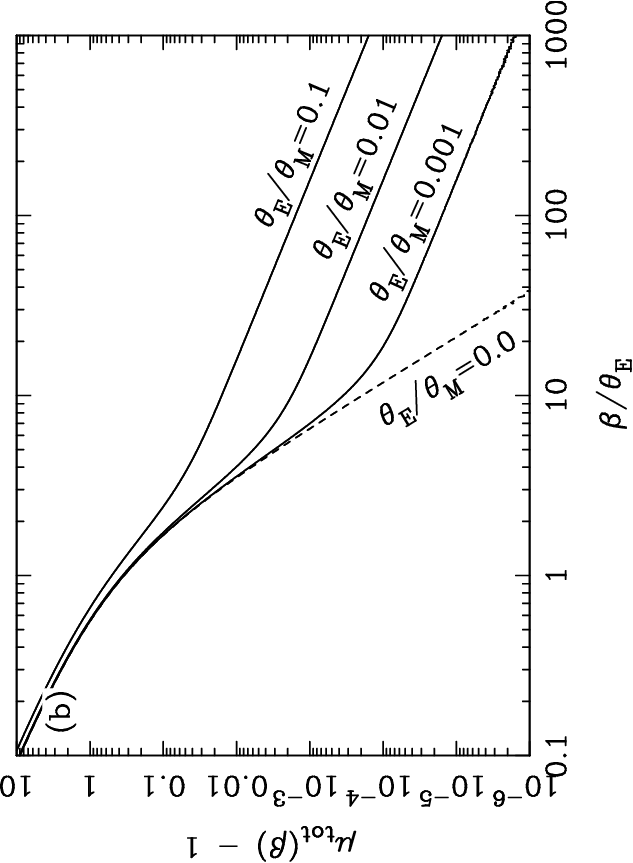}
\vspace{\doublefigureheight}
\caption{The total magnification of a source as a function of 
its angular separation, $\beta$, from the optical axis of a \mond ian 
\pointmass\ deflector. More extreme effects are emphasised in (a)
and weaker effects shown in (b). 
The solid curves show
increasing values of $\theta_{\rmn E} / \theta_{\rmn M}$,
as labelled;
the dashed curves show the Newtonian
result (which could also be obtained by taking 
$\theta_{\rmn E} / \theta_{\rmn M} = 0$). 
Note that $\theta_{\rmn E}$ is the Einstein angle of the Newtonian
\pointmass; the $x$-axis could also be normalised by the actual 
Einstein angle of the lens, which increases as $\theta_{\rmn M}$
decreases. This approach is taken in \fig{mu_ml}.}
\label{figure:mu_be}
\end{figure}

\begin{figure}
\includegraphics{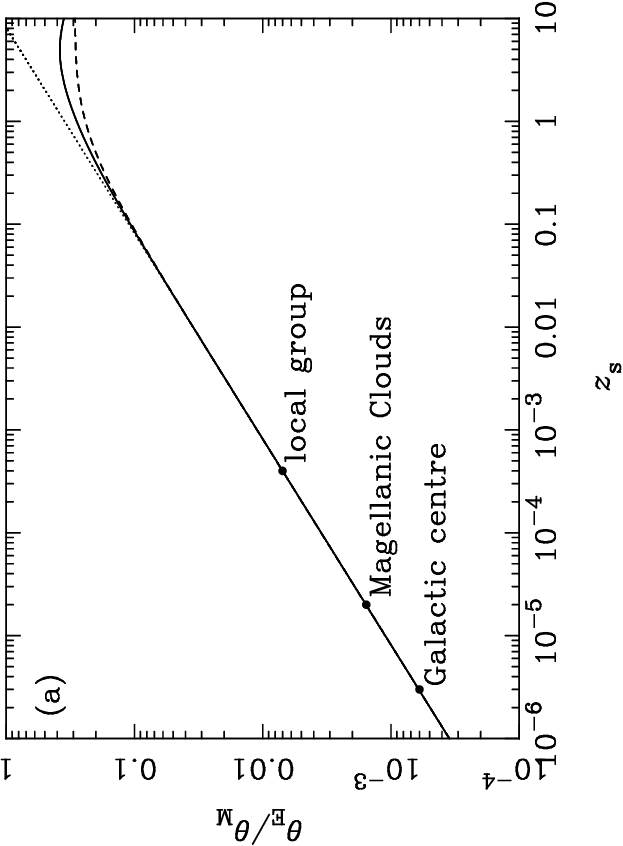}
\includegraphics{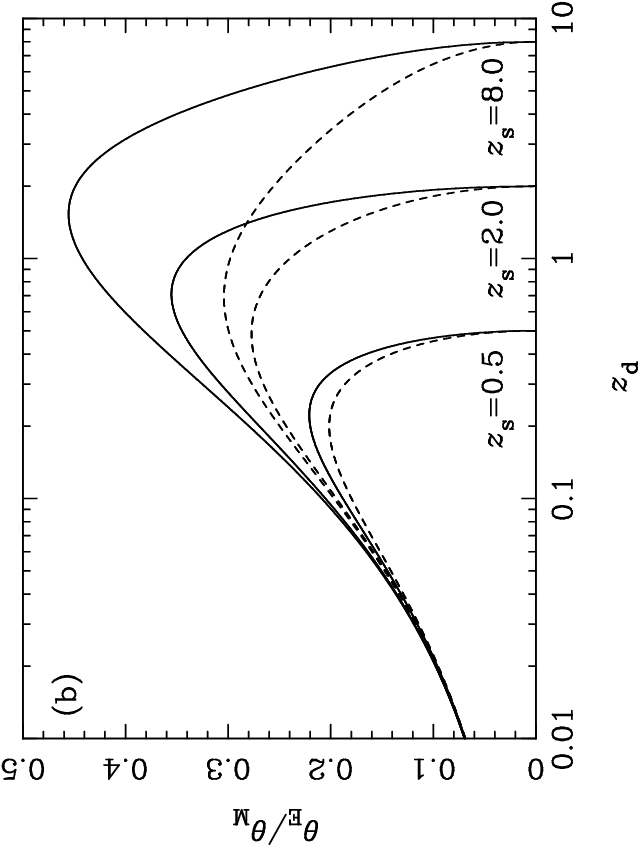}
\vspace{\doublefigureheight}
\caption{The ratio of the Einstein angle of a \pointmass, $\theta_{\rmn E}$,
to its characteristic
\mond ian angular scale, $\theta_{\rmn M}$, as a function of
of source redshift, $z_{\rmn s}$ (a) and
deflector redshift, $z_{\rmn d}$ (b).
In (a) the deflector is assumed to be halfway between observer and
source, and in (b) results are shown for several source redshifts,
as indicated.
Two cosmological models are shown:
$\omo = 0.01$ and $\olo = 0.99$ (solid lines);
and $\omo = 0.01$ and $\olo = 0.0$ (dashed lines);
the dotted line in (a) is given by
$\theta_{\rmn E}/\theta_{\rmn M} = 0.35 z_{\rmn s}^{1/2}$,
which is a good approximation in the local universe. Scales corresponding
to the Galactic centre, the Magellanic Clouds,
and the local group are marked,
with the `effective redshift' given by 
$z_{\rmn s} = H_0 / c \, d_{\rmn os}$.}
\label{figure:e/m}
\end{figure}

Inverting the lens equation [\eq{lensequation}]
and using the magnification formul\ae\ given in 
\eq{mu_be} and (\ref{equation:mu_point}) allows 
$\mu_{\rmn tot}(\beta)$ to be calculated.
This is shown in \fig{mu_be}, with the $x$-axis normalised to 
match the Newtonian Einstein radius.
In all cases $\mu_{\rmn tot}(\beta)$ matches the relativistic form
for $\beta \la \theta_{\rmn M}$; beyond that \mond\ always
results in a greater magnification. 
The shapes of these curves are determined by the mass-independent
ratio of the 
Einstein angle to the characteristic \mond ian angular scale.
From \eqs{theta_E} and (\ref{equation:theta_M}) this ratio is 
given by 
\begin{equation}
\label{equation:e/m}
\frac{\theta_{\rmn E}}{\theta_{\rmn M}}
= \sqrt{\frac{4 a_0}{c^2} \frac{d_{\rmn od} d_{\rmn ds}}{d_{\rmn os}}}
\la 0.006 \left(\frac{d_{\rmn os}}{1 {\rmn \,\, Mpc}} \right)^{1/2}
\simeq 0.35 z_{\rmn s}^{1/2},
\end{equation}
where the upper limit is calculated by assuming that 
$d_{\rmn od} \simeq d_{\rmn ds} \simeq d_{\rmn os} / 2$,
and the last expression is valid only in the local universe, 
where $d_{\rmn os} = c z_{\rmn s} / H_0$.
This ratio is shown in \fig{e/m}: 
its increase with $z_{\rmn s}$ is quite apparent, and
it is also important that it does not vary greatly with $z_{\rmn d}$
unless either $d_{\rmn od}$ or $d_{\rmn ds}$ is close to zero.
Within the Galaxy \mond ian lensing effects should be minimal,
with $\theta_{\rmn E}/\theta_{\rmn M} \la 0.001$, but 
they could become very important on cosmological scales,
a point which is explored in more detail in \sect{microlens}.

\subsubsection{Thin rod}
\label{section:rod}

In the standard theory of gravitational lensing, the 
deflection properties of most isolated mass distributions
depend only on their projected surface density. 
As argued in \sect{lightbend} the thin-lens approximation 
is unlikely to hold in \mond, and this can be illustrated 
by considering the lensing effect of a thin rod oriented along the \los.
Its density is given by 
\begin{equation}
\rho(\vect{r}) = 
\left\{
\begin{array}{lll}
0, & {\rmn if} & z < - l / 2, \\
& & \\
M / l \, \delta^2(\vect{R}), & {\rmn if} & - l / 2 \leq z \leq l / 2 \\
& & \\
0, & {\rmn if} & z > l / 2,
\end{array}
\right.
\end{equation}
where $M$ is the rod's mass and $l$ its length.
Inserting this definition into \eq{alpha_light} and evaluating the
required integrals gives the Newtonian acceleration as
\[
a_{{\rmn N}_x}
\!\! = \!
- \frac{G M}{l} \frac{x}{R^2}
\left[
\frac{z + l / 2}{\sqrt{R^2 + (z + l / 2)^2}}
-
\frac{z - l / 2}{\sqrt{R^2 + (z - l / 2)^2}} 
\right]
\!\!
,
\]
\[
a_{{\rmn N}_y}
\!\! = \! - \frac{G M}{l} \frac{y}{R^2}
\left[
\frac{z + l / 2}{\sqrt{R^2 + (z + l / 2)^2}}
-
\frac{z - l / 2}{\sqrt{R^2 + (z - l / 2)^2}} 
\right]
\!\!,
\]
\[
a_{{\rmn N}_z}
\!\! = \!
- \frac{G M}{l}
\left[
\frac{1}{\sqrt{R^2 + (z - l / 2)^2}} 
- 
\frac{1}{\sqrt{R^2 + (z + l / 2)^2}}
\right]
\!,
\]
\begin{equation}
\label{equation:alpha_rod}
\end{equation}
where, as before, $R^2 = x^2 + y^2$.
The \mond ian acceleration can be calculated directly from this
using \eq{mond},
but the results are cumbersome and so not presented in full. 

\begin{figure}
\includegraphics{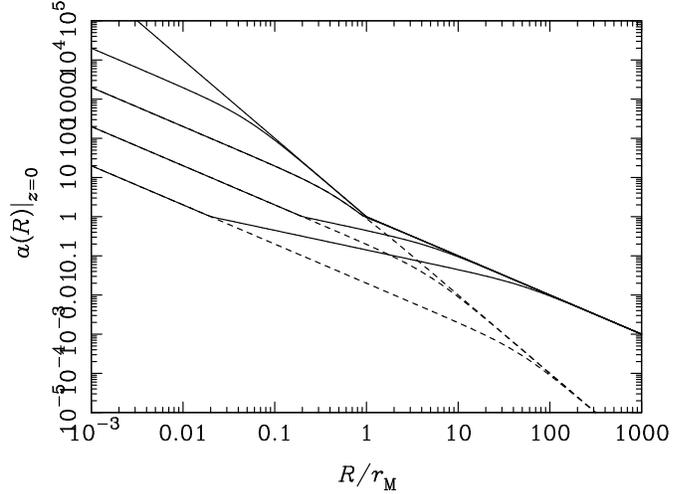}
\vspace{\singlefigureheight}
\caption{The equatorial acceleration produced by a thin rod 
(oriented along the $z$-axis)
in \mond\ (solid lines) and \gr\ (dashed lines).
Here $r_{\rmn M}$ is the radius at which the acceleration becomes
\mond ian for the corresponding \pointmass.
Curves are shown for several lengths:
0; 
$0.1 r_{\rmn M}$; 
$r_{\rmn M}$; 
$10 r_{\rmn M}$;
and $100 r_{\rmn M}$,
with the acceleration decreasing (at least for $r < r_{\rmn M}$) 
as length increases.}
\label{figure:acc_rod}
\end{figure}

Some idea of the lensing properties of such an object can
be gained by looking at the acceleration in the $x$-$y$ plane.
Applying the definitions given in \sect{model} yields an `equatorial'
\mond ian acceleration (directed towards the origin) of magnitude
\[
\left.a(R)\right|_{z = 0} = \left\{
\begin{array}{lll}
\frac{G M}{R}
\left[R^2 + (l / 2)^2\right]^{-1/2}, 
& \!\! {\rmn if} \!\! & R \leq R_{\rmn M}, \\
& & \\
\sqrt{\frac{G M a_0}{R}} 
\left[R^2 + (l / 2)^2\right]^{-1/4}, 
& \!\! {\rmn if} \!\! & R > R_{\rmn M} ,
\end{array}
\right.
\]
\begin{equation}
\end{equation}
where 
\begin{equation}
\label{equation:r_m_rod}
R_{\rmn M} = \frac{l}{2} \frac{1}{\sqrt{2}}
\left[
\sqrt{1 + 4 \left(\frac{r_{\rmn M}}{l / 2}\right)^4} - 1
\right]^{1/2}
\end{equation}
is the equatorial radius at which the acceleration equals $a_0$,
and $r_{\rmn M}$ is the \mond ian radius of the equivalent \pointmass.
A family of these curves is shown in \fig{acc_rod}, with the 
Newtonian accelerations also included for comparison.
Close to the rod $a(R) \propto R^{-1}$, the standard Newtonian result,
and for very large impact parameters $a(R) \propto R^{-1}$,
the rod acting like a \mond ian \pointmass. 
In most cases there is an intermediate region in which
the acceleration is sufficiently small to feel a \mond ian boost
but the rod's finite length is also important. The combination
of these two effects results in a $a(R) \propto R^{-1/2}$ dependence,
and any particle passing through this region would feel an unusually
large gravitational pull.

It is also revealing to consider how the acceleration 
varies along the $z$-axis. Taking $x = y = 0$ in \eq{alpha_rod}
and again applying \mond ian dynamics (outside the length of the rod)
yields 
\begin{equation}
\vect{a}(0, 0, z) = 
\end{equation}
\[
\mbox{}
\left\{
\!\!\!\!\!
\begin{array}{lll}
- \frac{G M}{l} \left(\frac{1}{|z| - l / 2} - \frac{1}{|z| + l / 2} \right)
\hat{\vect{z}},
& {\rmn if} & l/2 < |z| \leq z_{\rmn M}, \\
& & \\
- \sqrt{\frac{G M a_0}{l}} 
\left(\frac{1}{|z| - l / 2} - \frac{1}{|z| + l / 2} \right)^{1/2}
\hat{\vect{z}},
& {\rmn if} & |z| > z_{\rmn M}, 
\end{array}
\right.
\]
where 
\begin{equation}
\label{equation:z_m_rod}
z_{\rmn M} = \frac{l}{2} \sqrt{1 + \left( \frac{r_{\rmn M}}{l / 2} \right)^2}.
\end{equation}
Note that if $l \gg r_{\rmn M}$ the acceleration becomes \mond ian
for $|z| \ga l/2$, \ie\ anywhere outside the rod.

\begin{figure}
\includegraphics{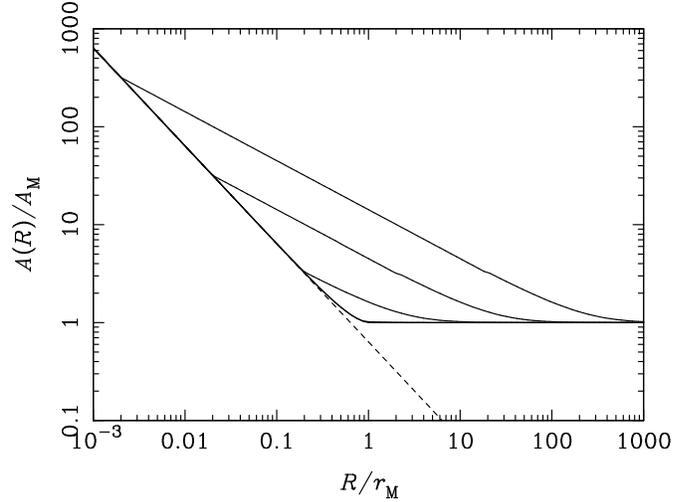}
\vspace{\singlefigureheight}
\caption{The deflection law of a thin rod oriented along the \los\
in \mond\ (solid lines) and \gr\ (dashed line).
Here $r_{\rmn M}$ is the radius at which the acceleration becomes
\mond ian for the corresponding \pointmass, and $A_{\rmn M}$ is the
asymptotic constant deflection angle.
Curves are shown for several lengths: 
0; 
$r_{\rmn M}$ (almost indistinguishable from $l = 0$); 
$10 r_{\rmn M}$; 
$100 r_{\rmn M}$; 
and $1000 r_{\rmn M}$,
with the deflection angle increasing with length.}
\label{figure:alpha_rod}
\end{figure}

The integrals required to obtain the deflection angle must be 
calculated numerically, and the results are shown in \fig{alpha_rod}.
If $R \ll R_{\rmn M}$ 
%[which is given in \eq{r_m_rod}] 
then 
the deflection is essentially Newtonian 
[\ie\ $A(R) \rightarrow 4 G M / (c^2 R)$] as the deflection is dominated
by the portion of the photon path alongside the rod. 
The opposite extreme is if $R \gg l$, in which case the rod acts 
like a \pointmass, being `unresolved' by the photon. Thus $A(R)
\rightarrow 2 \pi (G M a_0)^{1/2} / c^2 = A_{\rmn M}$ in this regime, provided
that the rod is of finite length\footnote{An infinitely long rod has an
infinite deflection angle, irrespective of impact parameter.}.
A corollary of these two arguments is that the thin-lens approximation
is valid in \mond\ if the \los\ extension of the deflector is 
less than $r_{\rmn M}$, as, for all impact parameters, the deflection
is either Newtonian or effectively point-like. 
However if $l \ga r_{\rmn M}$ then $A(R) \propto R^{-1/2}$ 
for $R_{\rmn M} \la R \la l$. In this case the photon's deflection
is dominated by a segment of length $\sim l$ during which it feels 
a nearly perpendicular acceleration of $a \simeq [2 G M a_0 / (R l)]^{1/2}$. 
Thus the deflection angle is $A(R) \simeq 2 / c^2 (2 G M a_0 l / R)^{1/2}$ 
in this region, and is proportional to the \los\ extent of the deflector.

The differences between the curves in \fig{alpha_rod}
give some idea of the the validity of the 
thin-lens approximation for an isolated deflector.
For example, a typical elliptical galaxy with its major axis along the \los\
would have an effective $l$ of $\sim 10$ kpc or more (\sect{galgal})
and $r_{\rmn M} \simeq 2$ kpc, which would imply that the thin-lens
approximation is probably reasonable in this situation.

% \subsubsection{Two \pointmasses}
% \label{section:twopoint}
% 
% Thus far rotationally invariant deflectors have been considered --
% this symmetry allows the direction of the photon deflection to 
% be assumed with confidence, but is clearly limited in its application.
% The simplest lens which breaks this symmetry consists of two 
% \pointmasses\ (assuming they are not aligned along the \los).
% The analysis of this model allows the degree to which the 
% direction of the deflection angle (as opposed to its magnitude)
% can vary for a given projected mass distribution.
% Without loss of generality, the masses are taken to lie in the 
% $x$-$z$ plane, giving
% \begin{eqnarray}
% \rho(\vect{r}) & = & 
% \frac{M}{2} \, \delta(x - \Delta x/2) \, \delta(y) \delta(z - \Delta z/ 2)
% \nonumber \\
% & + & 
% \frac{M}{2} \, \delta(x + \Delta x/2) \, \delta(y) \, 
% \delta(z + \Delta z/ 2), 
% \end{eqnarray}                                              
% where it is further assumed that they both have mass $M/2$.
% 
% \begin{figure}
% %\special{psfile="twopoint.ps" angle=-90 vscale=60 hscale=60
%       % %hoffset=-15 voffset=28}
% \vspace{\singlefigureheight}
% \caption{}
% \label{figure:twopoint}
% \end{figure}

\subsubsection{Isothermal sphere}
\label{section:iso}

The isothermal sphere (\eg\ Binney \& Tremaine 1987)
is a useful and simple galaxy model
within the \dm\ paradigm as it explains the flat rotation
curves of spirals, the dynamics of ellipticals, and the 
lensing properties of both. 
Parameterised by its (unobservable) velocity
dispersion, $\sigma$, and its core radius, $r_{\rmn c}$, its
density is given by (\eg\ Hinshaw \& Krauss 1987)
\begin{equation}
\rho_{\rmn N}(r) = \frac{\sigma^2}{2 \pi G} \frac{1}{R^2 + r_{\rmn c}^2}.
\end{equation}

Given the choice of $f_{\rmn M}(x)$ described in \sect{model}, the 
gravitational properties of this model can be almost 
exactly replicated in \mond\ by the density profile
\begin{equation}
\rho(r) = 
\left\{
\begin{array}{lll}
\frac{\sigma^2}{2 \pi G} \frac{1}{R^2 + r_{\rmn c}^2}, 
& {\rmn if} & r \leq r_{\rmn M}, \\
& & \\
0, & {\rmn if} & r > r_{\rmn M},
\end{array}
\right.
\end{equation}
assuming $r_{\rmn c} \ll r_{\rmn M}$, which is certainly the case
for real galaxies. 
The image positions and magnifications produced by such a mass distribution 
could be obtained from the \mond ian formalism (\sect{model}), 
but there is no need to perform these calculations
explicitly, as its lensing properties match
those of the conventional isothermal sphere, 
which has been studied in great detail 
(\eg\ Hinshaw \& Krauss 1987; Kochanek 1996; Mortlock \& Webster 2000).
In particular, standard calculations of lens statistics remain
valid, a point explored in more detail in \sect{n_lens}.

%\subsubsection{Ellipsoidal lenses}
%\label{section:ell}
%
%The typical deflectors that produce multiply-imaged quasars are 
%isolated elliptical galaxies (\eg\ ), which, if they contain no dark matter,
%have projected densities with elliptical symmetry and 
%spatial densities with ellipsoidal symmetry.
%There are several ways to define mass distributions that satisfy
%these symmetries, but homoeoidal symmetry (Chandrasekhar 19??)
%is chosen here due to the mathematical simplicity of the force calculation.
%Following Schramm (1990), 
%the density of the lens is given by $\rho(x, y, z) =
%\rho\left[ (x^2 + y^2 / b^2 + z^2 / c^2)^{1/2} \right]$,
%where $b$ and $c$ are the axial ratios in the $y$- and $z$-directions,
%respectively, relative to the $x$-direction.

\section{Galaxy-galaxy lensing}
\label{section:galgal}

Background galaxies are observed to be tangentially aligned
around foreground galaxies due to the latter
population's gravitational lensing effect.
Termed galaxy-galaxy lensing,
this technique has -- under the assumption of \gr\ --
confirmed the existence of extended
isothermal haloes
around all types of galaxies (\eg\ Brainerd \etal\ 1996; Fischer \etal\ 2000).
However it is potentially an even more 
powerful probe of alternative gravity theories as 
the shear signal extends far enough beyond the visible extent 
of the deflectors that they can be treated as point-masses.
The radial dependence of the
shear signal is consistent with the \mond ian lensing
formalism described in \sect{model} (Mortlock \& Turner 2001),
and the forthcoming 
Sloan Digital Sky Survey (\SDSS; York \etal\ 2000)
data should allow measurements out to physical separations of
$\sim 1$ Mpc. 
Beyond this the influence of secondary deflectors will start to dilute the
signal, but, if \mond\ returns to a Newtonian regime at ultra-low
accelerations ($\sim 10^{-12}$ m s$^{-2}$), the resultant cut-off might
be detectable.

The full \SDSS\ data-set will also facilitate a number of more 
general tests which should be able to discriminate between \dm\ 
and alternative gravity theories unambiguously. These are discussed
in detail by Mortlock \& Turner (2001), and so only summarised here. 
One implication of \mond\ is that the shear signal around any 
subset of the foreground population should have the same functional
form, within the errors; any departure from this would represent
strong evidence against such a model. 
A more powerful idea is to search for any deviation from
azimuthal symmetry in the lensing signal (\cf\ Natarajan \& Refregier 2000).
A positive detection would be difficult to reconcile with \mond\
(or any alternative gravity theory with basic symmetry properties);
a symmetric signal, on the other hand, would be in conflict
\dm-based halo formation models (\eg\ Navarro, Frenk \& White 1995, 1996).

\section{Microlensing}
\label{section:microlens}

Microlensing is the term used to describe the lensing action of 
(collections of) individual stars, and has been observed over
a wide range of scales, from within the Galactic halo to redshifts
of order unity.
As the name suggests, the resultant image separations are much 
smaller than present day telescopes can resolve, and its only
observable consequence is 
the change in magnification of a source caused by the 
relative motion of the 
deflector(s) across the \los. 

Given that the additive nature of \mond ian lensing is unknown, 
this discussion of microlensing will be limited to isolated 
deflectors, primarily single stars.
For a point-source of unlensed magnitude $m_0$, 
the resultant light-curve takes the form
(\cf\ Paczy\'{n}ski 1986)
\begin{eqnarray}
\label{equation:m(t)}
m(t) & = & m_0 \\
& - & \! 2.5 \log
\left\{ \! \mu_{\rmn tot}
\! \left[\beta_{\rmn 0}
\sqrt{
        \left(
        \frac{\beta_{\rmn min}}{\beta_{\rmn 0}}
        \right)^2
+ \left(\frac{t - t_{\rmn min}}{t_{\rmn 0}}\right)^2
}
\right] \!\!
\right\} 
, \nonumber
\end{eqnarray}
where 
$\mu_{\rmn tot}(\beta)$ is given in \sect{pointmass},
$\beta_{\rmn min} / \beta_0$ characterises the greatest
alignment of deflector and source,
$t_0$ is the time required for the alignment to change by angle $\beta_0$,
and
$t_{\rmn min}$ is the time at which the greatest alignment occurs.
The choice of $\beta_0$ is somewhat arbitrary, but it
is usual to define $\beta_0 = \theta_{\rmn E}$,
which implies that $\mu_{\rmn tot}(\beta_0) = 1.34$ in \gr. 
This last relation is taken to define $\beta_0$ here, 
which then implies that $\beta_0 \ge \theta_{\rmn E}$ in general,
with equality in the limiting case of 
$\theta_{\rmn M} \rightarrow \infty$.

\begin{figure}
\includegraphics{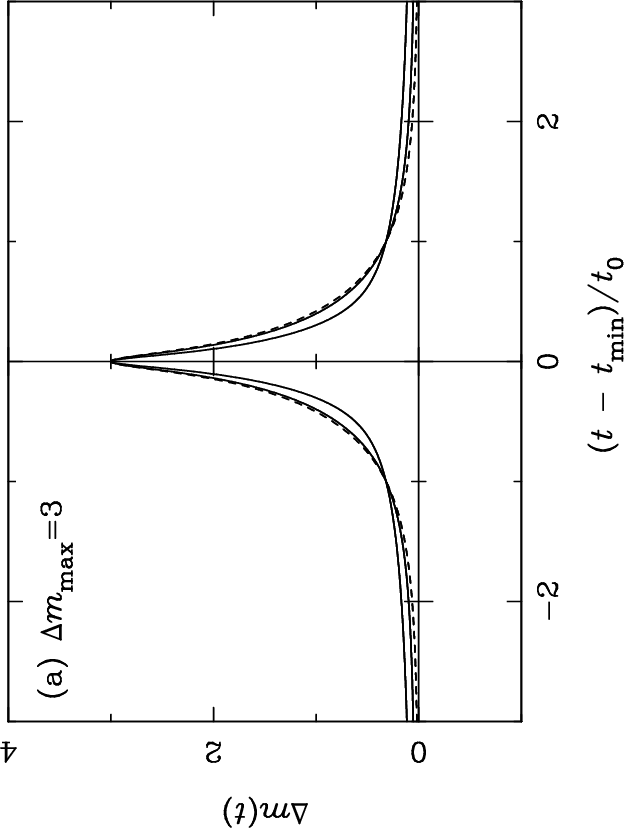}
\includegraphics{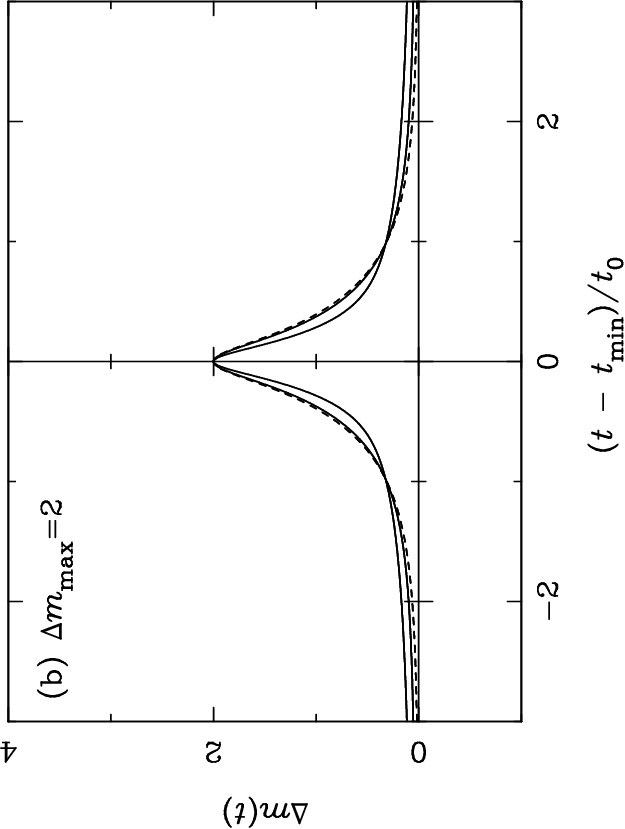}
\includegraphics{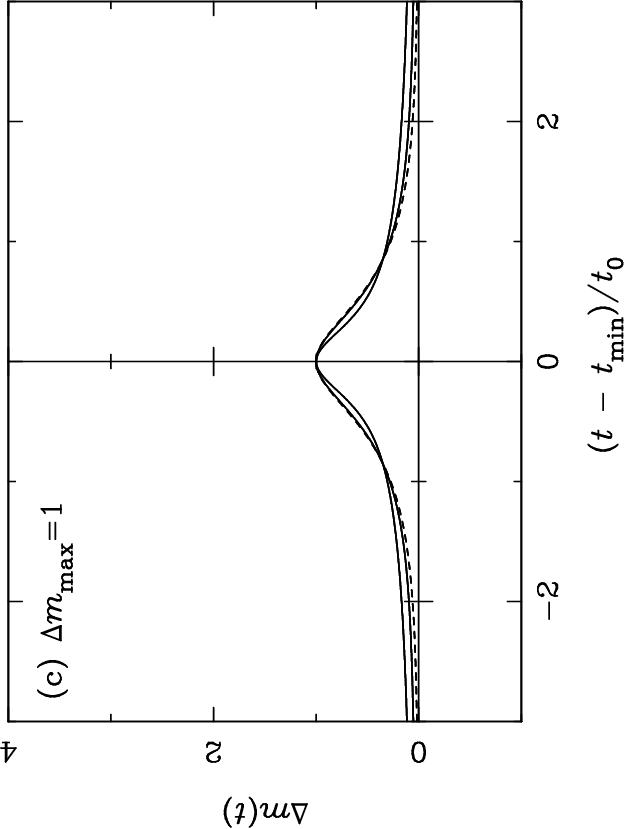}
\vspace{\triplefigureheight}
\caption{Microlensing light-curves,
$\Delta m(t) = m_0 - m(t)$, calculated assuming \mond\ (solid lines)
and \gr\ (dashed lines).
Each panel shows a different peak magnification (as labelled),
and in each case the Einstein radius (\ie\ the mass of the deflector)
and minimum impact parameter are chosen
to best fit the Newtonian result. 
For each value of $\Delta m_{\rmn max}$
light-curves are for 
$\theta_{\rmn E} / \theta_{\rmn M} = 0.0$ (\ie\ \gr),
$\theta_{\rmn E} / \theta_{\rmn M} = 0.1$ and 
$\theta_{\rmn E} / \theta_{\rmn M} = 1.0$.}
\label{figure:mu_ml}
\end{figure}

Several \mond ian microlensing
light-curves are shown in \fig{mu_ml}.
The distinction between
\mond\ and \gr\ is clear if
$\theta_{\rmn E} / \theta_{\rmn M} \ga 1$;
otherwise the only difference between the light-curves is in 
the $m(t) \simeq m_0$ wings.
The ratio $\theta_{\rmn E} / \theta_{\rmn M}$ 
depends mainly on the observer-source distance (\sect{pointmass}), and only
approaches unity if $z_{\rmn s} \ga 1$ (\fig{e/m}).
In physical terms, 
microlensing is useful as a probe of gravitational
theories if the magnification is significant
along geodesics passing beyond the
Newtonian region of the deflector.
This is not the case for microlensing within the local group
(\sect{galmicro}), but is true (in a \mond ian universe) on cosmological
scales (\sect{cosmicro}).

\subsection{Microlensing within the local group}
\label{section:galmicro}

The vast majority of discrete microlensing events observed to date
have been detected during the various monitoring programs 
(\eg\ Afonso \etal\ 1999;
Alcock \etal\ 2000; see Paczy\'{n}ski 1996 for a review)
directed at either 
the Galactic centre or the 
Magellanic Clouds,
for which $d_{\rmn os} \la 60$ kpc. 
Pixel lensing studies of M~31 (\eg\ Gyuk \& Crotts 2000; 
Cheongho \& Gould 1996) have extended the 
distance scale to $\sim 1$ Mpc, but this is still too nearby 
to be very sensitive to \mond ian effects.
\Eq{e/m} and \fig{e/m} imply
that $\theta_{\rmn E}/\theta_{\rmn M} \la 0.005$ 
for all microlensing events within the local group.
Basic lensing theory
(\eg\ Schneider \etal\ 1992) is then sufficient to show that only sources
magnified by less than one per cent would be subject to 
any additional \mond ian boost. To detect these changes, 
relative photometry accurate to $\sim 0.001$ mag would be 
required, and so
there is little chance of constraining alternative gravity 
theories from the data currently available on single light-curves. 

Future observations
of local group microlensing events might include very accurate photometric 
monitoring for long periods after the magnification peak, 
but the distinction between \mond ian and relativistic curves would 
only become apparent on scales where other deflectors along the \los\
start to have an effect.
One promising avenue of investigation is to `stack' the light-curves of 
simple microlensing events (\ie\ in which both the deflector and source 
are point-like); this is 
discussed in a more general context in Mortlock \& Turner (2001).
Such an approach would not only allow a distinction to be made 
between \gr\ and \mond, but it would also facilitate a 
measurement of $f_{\rmn M}(x)$ (defined in \sect{model}).

\subsection{Cosmological microlensing}
\label{section:cosmicro}

For sources at redshifts of $\ga 1$ 
the Einstein radius and \mond\ radius of a \pointmass\ deflector 
become comparable [\eq{e/m} and \fig{e/m}].
The resultant microlensing light-curves are then quite different,
as illustrated in \fig{mu_ml}, and there is the possibility of 
a clean and simple test to distinguish between \gr,
\mond, and other gravitational theories (Mortlock \& Turner 2001).

In order to test these predictions, accurate photometry of 
cosmological microlensing by an isolated deflector is required. 
Such data could be forthcoming from the quasar monitoring 
programs being undertaken by Walker (1999) and Tadros, Warren \& Hewett (2001).
The idea of both experiments is to observe quasars
seen through galactic or cluster haloes in the hope of a positive
detection of compact \dm. If \mond\ is correct then many of these
targets should show no signs of microlensing, as they 
lie along empty (\ie\ dark) \loss, which should be devoid of 
any potential deflectors in this model (\eg\ Walker 1994;
Mortlock \& Turner 2001). 
It is thus quite likely that 
serendipitous observations of microlensing may 
provide the desired data,
and indeed the spurious peak in 
the light-curve of gamma ray burst (\grb) 000301C 
(Sagar \etal\ 2000, and references therein)
may be the result of microlensing (Garnavich, Loeb \& Stanek 2000).
Unfortunately the uncertainties in the photometry and
lack of knowledge of the source geometry prevent any 
discrimination between \mond\ and \gr\ (Mortlock \& Turner 2001).
Nonetheless, more events should be forthcoming, 
and cosmological microlensing has the potential to be a very
clean method by which to measure the deflection law of a \pointmass.

\section{Quasar lensing}
\label{section:quasars}

Since the identification of the first gravitationally-lensed object -- 
Q~0957+561 (Walsh, Carswell \& Weymann 1979) -- there have been
more than fifty quasar lenses discovered. 
The image configurations have been used to map the \dm\ in the lens 
galaxies (\eg\ Chen, Kochanek \& Hewitt 1995;
Chae, Turnshek \& Khersonsky 1998; Keeton, Kochanek \& Falco 1998), 
and the fraction of quasars which are lensed places
strong limits on the cosmological model (\eg\ Kochanek 1996).
In the context of a \mond ian universe, individual lenses can be
used to constrain the deflection law of an extended mass distribution
(\sect{quasarmodel}) and the lens statistics are similarly
revealing in a global context (\sect{n_lens}).

\subsection{Lens modeling}
\label{section:quasarmodel}

If there is no \dm\ then it must be possible to model any quasar lens
with a mass distribution that, when integrated along the \los, also
matches the observed surface brightness of the deflector.
Other mass concentrations along the \los\ may also have some 
influence (\eg\ Keeton, Kochanek \& Seljak 1997), but this is 
potentially a very powerful test of \mond.

The fact that the three-dimensional mass distribution is required 
to perform such a test suggests elliptical galaxies as 
deflectors,
and several cases of lensing by
(apparently) isolated ellipticals have already been discovered: \eg\ 
MG~1654+1346 (Langston \etal\ 1989);
MG~1549+3047 (Leh\'{a}r \etal\ 1993);
HST~1253$-$2914 (Ratnatunga \etal\ 1995); 
and
HST~14176+5226 (Ratnatunga \etal\ 1995).
Assuming that the deflector is a triaxial ellipsoid,
the constraints implied by the observed surface brightness distribution
leave only one free parameter, which is essentially the unknown
extension of the galaxy along the \los. 
This is sufficiently restrictive that a successful fit would probably
represent positive evidence in favour of \mond.
However preliminary investigations suggest that such
modeling cannot always produce a good fit to both photometric
and lensing data, which would pose 
serious problems for the \mond ian lensing theory put forward in 
\sect{model}. 

\subsection{Lens statistics}
\label{section:n_lens}

The fraction of high-redshift sources that are multiply-imaged
increases rapidly with the cosmological constant as the co-moving
volume element in an $\olo$-dominated universe is so high
(Turner 1990). Kochanek (1996) used a selection of optical data
and sophisticated modeling to show that the low observed number
of lenses implies $\olo < 0.65$ (with 95 per cent confidence),
a limit which is only marginally consistent with the 
conventional \dm-dominated universes
preferred by \cmb\ data and high-redshift supernova observations 
(\eg\ Efstathiou \etal\ 1999).
As a \mond ian universe is likely to obey the Friedmann equations
(Sanders 1998), these lensing results are completely at odds
with the $\omo \simeq 0.01$, $\olo \simeq 0.99$ model 
described in \sect{intro}.
With the matter density limited by nucleosynthesis, 
the universe must be dominated by the vacuum energy density
in order to satisfy the \cmb\ power spectrum (McGaugh 2000).
Thus it seems that no single cosmological model can explain 
lens statistics, supernova data and \cmb\ anisotropies within
the framework of \mond.

One possible flaw in the above arguments is the
assumed form of the angular diameter distance, 
which is not known with any certainty in \mond. 
The lens statistics are mildly sensitive to the distance 
formula used (Ehlers \& Schneider 1986); 
the main aim of the supernova observations is to measure 
the luminosity distance; 
and the angular scale of the first peak in the \cmb\ power spectrum
only implies a flat universe if the angular 
diameter distance (on degree scales) obeys the standard formul\ae\
(\eg\ Carroll \etal\ 1992).
Despite this formal ambiguity, the angular diameter distance would
have to differ quite radically from their form in \gr\ to prevent 
the low number of quasar lenses observed 
being a serious problem for \mond.

\section{Cluster lensing}
\label{section:cluster}

The central regions of galaxy clusters are
very effective gravitational lenses, as evidenced by 
giant arcs, the strongly distorted images of background
galaxies (\eg\ Kneib \etal\ 1996). 
Combined with the more numerous weakly sheared images of background
sources, the projected mass distributions
of lensing clusters have
been convincingly reconstructed in a number of cases 
(\eg\ Mellier \etal\ 1993; Hammer \etal\ 1997; 
AbdelSalam \etal\ 1998). 
The inferred mass maps are generally correlated with
the distribution of galaxies within the clusters,
but there is also strong evidence for \dm.

In \mond, on the other hand, the lensing signal must be solely
due to the cluster members (and any gas present). 
Just as lensing by isolated galaxies offers a powerful test of
\mond\ (\sect{quasarmodel}), so does cluster lensing.
The correlation between the mass maps and the cluster galaxies 
is certainly consistent with expectations, but the presence of 
a large number of deflectors makes a more quantitative analysis
problematic.
Most fundamentally, the net effect of multiple lenses is not known
with any certainty, despite being well-defined in the particular
formalism described in \sect{model}.
Further, even if the lens theory was known,
the lack of a thin-lens approximation results in a mass of 
degeneracies, with the \los\ location of each cluster galaxy 
being a free parameter. 
In some cases the discrepancy between theory and observation 
might be so severe that this ambiguity is unimportant (\eg\ 
Beckenstein \& Sanders 1994), but for
the moment no strong conclusions regarding 
\mond\ can be drawn from cluster lensing. 

\section{Conclusions}
\label{section:concs}

\mond\ is a modification of inertia (or gravity) which can explain a 
variety of astronomical 
observations that would otherwise imply the 
existence of large amounts of \dm. 
Lacking a relativistic analogue, however, \mond\ makes no definite 
predictions about cosmology or photon propagation.
Sanders (1998) took an empirical approach to the former problem
with some success, and, in a similar spirit, Mortlock \& Turner (2001) 
showed that the existing galaxy-galaxy lensing data are consistent with 
the simple \mond ian theory of gravitational lensing 
described by Qin \etal\ (1995).
This formalism was extended in 
\sect{model}, and can be applied to isolated mass distributions 
with some confidence, but, due to the counter-intuitive nature of \mond,
cannot be extrapolated to multiple deflectors with any certainty.
A further complication is the failure of the thin-lens approximation,
which means that any ambiguities in the (three-dimensional)
luminosity density of a deflector flow through to its lensing 
properties.
These difficulties notwithstanding, 
this tentative theory of \mond ian gravitational lensing 
is subject to a number of observational tests.

\mond\ is consistent with observations of galaxy-galaxy
lensing, although tests for deviation from azimuthal symmetry
in the shear signal should be able to discriminate unambiguously between
\dm\ and most alternative gravity theories 
(\sect{galgal}; Mortlock \& Turner 2001).
Low-optical depth microlensing within the local group 
is unlikely to be a particularly useful in this context,
but on cosmological scales will be a very clean probe of \mond,
if sufficiently many lensing events are detected (\sect{microlens}).
If \mond\ is consistent with such simple lensing
observations, more complex scenarios, such as strong lensing
by galaxies and clusters, should provide additional constraints
(\sects{quasars} and \ref{section:cluster}, respectively). 
Finally, the frequency of strong lensing events is already 
at odds with \mond,
as the $\omo \simeq 0.01$, $\olo \simeq 0.99$ 
\mond ian cosmology implied by \cmb\ observations (McGaugh 2000) 
should result in many more multiply-imaged sources than are observed
(\sect{n_lens}),
although there is some ambiguity in the redshift-distance relationship
in such a universe.

On balance the formalism described in \sect{model}
is a reasonable hypothesis for gravitational lensing 
within the framework for \mond, and must be 
qualitatively correct for isolated deflectors. 
The theory is subject to a number of tests,
although most await the completion of current 
surveys or further theoretical development.
However it is likely that any fully relativistic 
extension of \mond\ must be completely non-linear to 
explain all of the above manifestations of 
gravitational lensing.

\section*{Acknowledgments}

The authors acknowledge a number of 
interesting discussions with
Anthony Challinor,
Erwin de Blok,
Mike Hobson,
Geraint Lewis,
Stacy McGaugh,
Moti Milgrom,
Bohdan Paczy\'{n}ski,
Joachim Wambsganss.
DJM was funded by PPARC and 
this work was supported in part by NSF grant AST98-02802.

\bsp
\label{lastpage}
\end{document}

%% file: mnfonts.tex
\newif\ifAMStwofonts
\ifoldfss
  \newcommand{\rmn}[1] {{\rm #1}}

  \ifCUPmtlplainloaded \else
    \NewTextAlphabet{textbfit} {cmbxti10} {}
    \NewTextAlphabet{textbfss} {cmssbx10} {}
    \NewMathAlphabet{mathbfit} {cmbxti10} {} % for math mode
    \NewMathAlphabet{mathbfss} {cmssbx10} {} %  "   "    "
  \fi
  \ifAMStwofonts
    \ifCUPmtlplainloaded \else
      \NewSymbolFont{upmath} {eurm10}
      \NewSymbolFont{AMSa} {msam10}
      \NewMathSymbol{\upi}     {0}{upmath}{19}
      \NewMathSymbol{\umu}     {0}{upmath}{16}
      \NewMathSymbol{\upartial}{0}{upmath}{40}
      \NewMathSymbol{\leqslant}{3}{AMSa}{36}
      \NewMathSymbol{\geqslant}{3}{AMSa}{3E}

      \let\leq=\leqslant 
      \let\geq=\geqslant \let\ge=\geqslant
    \fi
  \fi
\fi % End of OFSS

\ifnfssone
  \newmathalphabet{\mathit}
  \addtoversion{normal}{\mathit}{cmr}{m}{it}
  \addtoversion{bold}{\mathit}{cmr}{bx}{it}
  \newcommand{\rmn}[1] {\mathrm{#1}}

  \newmathalphabet{\mathbfit} % math mode version of \textbfit{..}
  \addtoversion{normal}{\mathbfit}{cmr}{bx}{it}
  \addtoversion{bold}{\mathbfit}{cmr}{bx}{it}
  \newmathalphabet{\mathbfss} % math mode version of \textbfss{..}
  \addtoversion{normal}{\mathbfss}{cmss}{bx}{n}
  \addtoversion{bold}{\mathbfss}{cmss}{bx}{n}
  \ifAMStwofonts
    \ifCUPmtlplainloaded \else
      \UseAMStwoboldmath
      \makeatletter
      \new@mathgroup\upmath@group
      \define@mathgroup\mv@normal\upmath@group{eur}{m}{n}
      \define@mathgroup\mv@bold\upmath@group{eur}{b}{n}
      \edef\UPM{\hexnumber\upmath@group}
      \new@mathgroup\amsa@group
      \define@mathgroup\mv@normal\amsa@group{msa}{m}{n}
      \define@mathgroup\mv@bold\amsa@group{msa}{m}{n}
      \edef\AMSa{\hexnumber\amsa@group}
      \makeatother
      \mathchardef\upi="0\UPM19
      \mathchardef\umu="0\UPM16
      \mathchardef\upartial="0\UPM40
      \mathchardef\leqslant="3\AMSa36
      \mathchardef\geqslant="3\AMSa3E

      \let\leq=\leqslant 
      \let\geq=\geqslant \let\ge=\geqslant
    \fi
  \fi
\fi % End of NFSS release 1

\ifnfsstwo
  \newcommand{\rmn}[1] {\mathrm{#1}}

  \DeclareMathAlphabet{\mathbfit}{OT1}{cmr}{bx}{it}
  \SetMathAlphabet\mathbfit{bold}{OT1}{cmr}{bx}{it}
  \DeclareMathAlphabet{\mathbfss}{OT1}{cmss}{bx}{n}
  \SetMathAlphabet\mathbfss{bold}{OT1}{cmss}{bx}{n}
  \ifAMStwofonts
    \ifCUPmtlplainloaded \else
      \DeclareSymbolFont{UPM}{U}{eur}{m}{n}
      \SetSymbolFont{UPM}{bold}{U}{eur}{b}{n}
      \DeclareSymbolFont{AMSa}{U}{msa}{m}{n}
      \DeclareMathSymbol{\upi}{0}{UPM}{"19}
      \DeclareMathSymbol{\umu}{0}{UPM}{"16}
      \DeclareMathSymbol{\upartial}{0}{UPM}{"40}
      \DeclareMathSymbol{\leqslant}{3}{AMSa}{"36}
      \DeclareMathSymbol{\geqslant}{3}{AMSa}{"3E}

      \let\leq=\leqslant 
      \let\geq=\geqslant \let\ge=\geqslant
    \fi
  \fi
\fi % End of NFSS release 2

\ifCUPmtlplainloaded \else
  \ifAMStwofonts \else % If no AMS fonts
    \def\upi{\pi}
    \def\umu{\mu}
    \def\upartial{\partial}
  \fi
\fi